\documentclass[%
  reprint,
  amsmath,amssymb,
  aps,prb,
  superscriptaddress,
  longbibliography,
  floatfix,
]{revtex4-2}

\usepackage{graphicx}
\graphicspath{{figures_png_eps/}{./}}
\usepackage{dcolumn}
\usepackage{booktabs}
\usepackage{array}
\usepackage{bm}
\usepackage{hyperref}
\hypersetup{hidelinks}
\usepackage{xcolor}
\usepackage{orcidlink}

\newcommand{\bav}{\langle\beta\rangle}
\newcommand{\bavi}{\langle\beta_i\rangle}
\newcommand{\expq}{\mathrm{exp}_q}
\newcommand{\KGO}{\textsc{KGO}}
\newcommand{\DO}{\textsc{DO}}
\newcommand{\kB}{k_{\mathrm{B}}}
\newcommand{\rd}{\mathrm{d}}
\newcommand{\epsn}{\varepsilon_n}

\begin{document}

\title{Hyperstatistical thermodynamics of the one-dimensional
       Klein--Gordon and Dirac oscillators:\\
       a closed-form $q$-generalized Boltzmann factor and a
       quantitative comparison with Beck's superstatistics}

\author{A.~Boumali\,\orcidlink{0000-0003-2552-0427}}
\email[Corresponding author: ]{boumali.abdelmalek@gmail.com}
\affiliation{Laboratory of Theoretical and Applied Physics,
             Echahid Cheikh Larbi Tebessi University,
             12000 Tebessa, Algeria}

\date{\today}

% =================================================================
\begin{abstract}
We investigate the thermodynamics of the one-dimensional Klein--Gordon
oscillator (\KGO) and the one-dimensional Dirac oscillator (\DO)
within two closely related generalized-statistical frameworks: Beck's
superstatistics, used in its low-energy asymptotic form, and the
recently proposed hyperstatistics of Squillante, Soares, Tsallis and
de Souza.  The objective is to determine which framework gives the most
controlled and physically consistent description of relativistic
oscillator thermodynamics when the ordinary Boltzmann factor is replaced
by a non-Boltzmann statistical weight.  In hyperstatistics, a
$\gamma$-distribution of Boltzmann factors at the level of domains of
the system leads, after the internal Laplace transform and the domain
average, to the effective closed-form generalized Boltzmann factor
$B_q(\varepsilon)=\expq(-\bav\varepsilon)$.  We compute the partition
function, entropy and specific heat using excitation energies
$\epsn=E_n-E_0$, which remove the irrelevant rest-energy offset and
ensure the third-law behaviour $C_v\to0$ as $1/\bav\to0$.  The proper
one-dimensional degeneracies, $g_n=1$ for the \KGO\ and
$g_0=1$, $g_n=2$ for $n\ge1$ for the \DO, are retained throughout.
The results show that hyperstatistics reproduces the Boltzmann limit,
remains positive and smooth over the whole numerical range considered,
and distinguishes the \KGO\ from the \DO\ through the entropy increase
and the specific-heat shift produced by the spin-induced degeneracy of
the Dirac spectrum.  By contrast, Beck's truncated polynomial expansion
agrees with hyperstatistics only for $q-1\ll1$ and small
$\bav\varepsilon$; outside this regime, the Gamma and Log-Normal
brackets may become negative, whereas the F-distribution bracket may
grow unphysically.  We conclude that hyperstatistics is the most robust
working framework for the present thermodynamic sums, while exact
superstatistics remains preferable when the microscopic distribution of
intensive variables is known and the full Laplace transform can be used.
\end{abstract}


\maketitle

% =================================================================
\section{Introduction}
\label{sec:intro}
% =================================================================

The Klein--Gordon and Dirac oscillators are the canonical
relativistic generalizations of the harmonic oscillator and have
been at the centre of a broad range of theoretical and
experimental investigations.  The Klein--Gordon
oscillator~(\KGO)~\cite{Bruce1993,Boumali2015} is obtained by the
non-minimal substitution
$\hat{p}\to \hat{p}-\mathrm{i}m_0\omega\hat{x}$ in the
Klein--Gordon equation; the Dirac oscillator~(\DO),
introduced by Moshinsky and
Szczepaniak~\cite{Moshinsky1989} after earlier related work by
It\^o, Mori and Carriere~\cite{Ito1967}, is built from the same
substitution applied to the Dirac equation, in which case it
reduces in the non-relativistic limit to a harmonic oscillator
with a strong spin-orbit coupling term.  Both models are exactly
solvable, share an energy spectrum that grows as $\sqrt{n}$ rather
than $n$, and bridge non-relativistic harmonic motion and
relativistic quantum mechanics in a particularly clean way.  The
formal mapping of the \DO\ onto the Jaynes--Cummings model of
quantum optics~\cite{Bermudez2007} has motivated numerous
applications and analogue realizations, ranging from quantum
simulation with trapped ions~\cite{Lamata2011,Blatt2012} to
graphene-like Dirac materials~\cite{Boumali2015graphene}, and the
1D~\DO\ has even been realised experimentally in a microwave
billiard~\cite{Franco2013}.

When such a system is embedded in a non-equilibrium environment,
or in a thermal bath whose intensive parameters fluctuate slowly
on the system's microscopic time-scale, the canonical Boltzmann
factor $\mathrm{e}^{-\beta E}$ no longer captures its statistical
properties.  Beck and Cohen~\cite{Beck2003} introduced
\emph{superstatistics} to cope with this situation by writing the
generalized Boltzmann factor as the Laplace transform of a
probability density $f(\beta)$ over the inverse temperature,
\begin{equation}
B(E) \;=\; \int_0^{\infty}\! f(\beta)\,\mathrm{e}^{-\beta E}\,\rd\beta,
\label{eq:BcohenLT}
\end{equation}
and showed that, for small $\bav E$, $B(E)$ is universal in the
parameters $q\equiv\langle\beta^2\rangle/\bav^2$ and $\bav$, with a
leading correction that is polynomial in $\bav E$ and a third-order
coefficient $g(q)$ that depends on the chosen $f(\beta)$.  The
three most popular choices --- Gamma, Log-Normal and
F-distribution --- give~\cite{Beck2003,Beck2011}
\begin{align}
g_{\Gamma}(q)        &= -\tfrac{1}{3}(q-1)^2,                          \label{eq:gGamma}\\
g_{\mathrm{LogN}}(q) &= -\tfrac{1}{6}\bigl(q^3-3q+2\bigr),             \label{eq:gLogN}\\
g_{F}(q)             &= -\tfrac{1}{3}\,\frac{(q-1)(5q-6)}{3-q}.        \label{eq:gF}
\end{align}
The asymptotic expansion behind Eqs.~\eqref{eq:gGamma}--\eqref{eq:gF}
is valid only in a window of $\bav E$ small enough that the
polynomial bracket
\begin{equation}
\Pi_q(\bav E) \;\equiv\; 1+\tfrac{q-1}{2}(\bav E)^2 + g(q)(\bav E)^3
\label{eq:Pi_def}
\end{equation}
remains positive and monotonically decreasing.  As we will see in
Sec.~\ref{sec:results}, this bracket loses positivity at moderate
$\bav E$ for the Gamma and Log-Normal PDFs at $q\gtrsim 1.2$, and
diverges upward for the F-distribution at $q\gtrsim 1.1$ already at
$\bav\varepsilon\approx 4$, behaviours that contradict the very interpretation
of $B(E)$ as a Boltzmann-like weight.  This is the well-recognised
practical limitation of asymptotic superstatistics that the present
analysis brings into sharp relief.

The non-additive Tsallis $q$-entropy~\cite{Tsallis1988},
\begin{equation}
S_q \;=\; \kB\,\frac{1-\sum_i p_i^{q}}{q-1},
\qquad
S_1=-\kB\sum_i p_i\ln p_i,
\label{eq:Sq}
\end{equation}
recovers Boltzmann--Gibbs statistics in the limit $q\to 1$ and has
proven indispensable in describing complex systems with long-range
interactions, multifractal structure, or heavy-tailed
distributions.  Beck and Cohen showed that superstatistics with a
$\gamma$-distribution of $\beta$ leads asymptotically to
$q$-exponential weights, identifying superstatistics as a dynamical
foundation for nonextensive statistical
mechanics~\cite{Beck2003,Beck2011}.  However, the connection is
asymptotic: the $q$-exponential is recovered only at small
$\bav E$, and the choice of $f(\beta)$ enters Beck's framework
explicitly through the polynomial coefficient $g(q)$.

Squillante, Soares, Tsallis and de~Souza~\cite{Squillante2026} have
recently proposed \emph{hyperstatistics}, in which the
$\gamma$-distribution is taken at the level of \emph{Boltzmann
factors inside domains} of the system rather than at the level of
intensive parameters.  The Laplace transform of the
$\gamma$-distribution then yields the $q$-exponential exactly, and
a subsequent average over any normalisable $f(\beta)$ returns
\begin{equation}
B_q(E) \;=\; \int_0^{\infty}\!\rd\beta_i\, f(\beta_i)\,
        \expq\!\bigl(-\bavi E\bigr) \;=\; \expq\!\bigl(-\bav E\bigr),
\label{eq:Bq}
\end{equation}
where the $q$-exponential is
\begin{equation}
\expq(x) \;=\; \bigl[1+(1-q)x\bigr]^{1/(1-q)},
\qquad
\mathrm{exp}_1(x) = \mathrm{e}^{x}.
\label{eq:qexp}
\end{equation}
Equation~\eqref{eq:Bq} is the central result of
Ref.~\cite{Squillante2026}: the closed $q$-exponential form of
$B_q(E)$ persists for Uniform, $\gamma$, Log-Normal, F and the
newly introduced $q$-$\gamma$ probability densities, with only the
prefactor in the argument depending on $f(\beta)$.  Following the
clarification received from one of the authors of
Ref.~\cite{Squillante2026}, Eq.~\eqref{eq:Bq} should be read as a
domain-averaged effective construction: after the internal
$\gamma$-transform has produced $\expq(-\bavi E)$ in each domain,
the domain mean is replaced by the measured system mean $\bav$.
It is not a general Jensen-type identity of the form
$\int f(\beta)\expq(-\beta E)\,\rd\beta=\expq(-\bav E)$,
which would be false for a nonlinear $q$-exponential.  With this
interpretation, hyperstatistics is a remarkable structural
simplification compared with asymptotic superstatistics, since (i)
the working weight is a closed $q$-exponential rather than a finite
polynomial, (ii) the final form is not selected case by case by the
choice of $f(\beta)$, and (iii) it propagates analytically into all
thermodynamic relations through the simple identity
$\partial_{\bav}\expq(-\bav E)=-E[\expq(-\bav E)]^{q}$.

The thermodynamic properties of the \KGO\ and \DO\ have been
investigated extensively in recent years.  Pacheco, Landim and
Almeida~\cite{Pacheco2003} computed the partition function and
specific heat of the 1D~\DO\ in a thermal bath; the corresponding
3D analysis was given by Pacheco, Maluf, Almeida and
Landim~\cite{Pacheco2014}.  The relativistic harmonic oscillators
in 1D have been studied analytically through the Hurwitz zeta
function and Epstein--Riemann methods by
Boumali~\cite{Boumali2015,Boumali2015graphene,Boumali2018DOmin},
and by Boumali and collaborators in noncommutative
spaces~\cite{Boumali2013}, in cosmic-string
backgrounds~\cite{Bouzenada2024}, in fractional
spaces~\cite{Korichi2021}, and most recently in the doubly special
relativity (DSR) frameworks of Amelino--Camelia and
Magueijo--Smolin~\cite{Boumali2025DSR,Boumali2026KG3D}.  The
behaviour of these systems under Beck's asymptotic superstatistics
has also been analysed: the 1D, 2D and 3D~\KGO\ in
Ref.~\cite{Siouane2024JLTP}, and the 1D~\DO\ with the same Gamma,
Log-Normal and F prescriptions in Ref.~\cite{Siouane2024TMP}.

The central objective of the present paper is to assess, within a
single and analytically transparent relativistic-oscillator setting,
whether hyperstatistics provides a more controlled thermodynamic
framework than the asymptotic form of Beck's superstatistics.  This
objective is pursued through four specific steps:
\begin{enumerate}
\item we reformulate the asymptotic-superstatistics analysis of
      Ref.~\cite{Siouane2024JLTP} for the 1D \KGO\ in a notation
      directly comparable with hyperstatistics;
\item we extend the same comparison to the 1D \DO, retaining the
      degeneracy structure imposed by the spectrum of
      Pacheco \emph{et al.}~\cite{Pacheco2003};
\item we compute the entropy and specific heat of both oscillators
      with the closed hyperstatistical weight of
      Eq.~\eqref{eq:Bq} and compare them with the Gamma,
      Log-Normal and F versions of Beck's asymptotic expansion;
\item we identify the parameter range in which Beck's polynomial
      bracket remains a positive Boltzmann-like weight and determine
      how its breakdown affects the thermodynamic functions.
\end{enumerate}
This structure makes the comparison quantitative rather than purely
qualitative: the same spectra, degeneracies and excitation energies
are used in both approaches, so that any difference in the curves can
be traced directly to the statistical weight.

The paper is organized as follows.  Section~\ref{sec:spectrum}
collects the spectrum and degeneracies of the 1D~\KGO\ and
1D~\DO.  Section~\ref{sec:beck} reformulates Beck's low-energy
superstatistics for these systems.  Section~\ref{sec:hyper}
develops the hyperstatistical framework and writes the partition
function in closed form.  Section~\ref{sec:thermo} states the
thermodynamic prescription common to both frameworks and lists the
explicit analytical formulas for the hyperstatistical case.
Section~\ref{sec:results} presents and discusses the numerical
results through six figures.  Section~\ref{sec:conclusion}
concludes.

% =================================================================
\section{Spectrum and degeneracies in one dimension}
\label{sec:spectrum}
% =================================================================

\subsection{The 1D Klein--Gordon oscillator}
\label{subsec:KGO}

The 1D~\KGO\ is obtained from the Klein--Gordon equation through
the non-minimal coupling
$p\to p-\mathrm{i}m_0\omega\hat{x}$ in a single component, leading
to the eigenvalue
problem~\cite{Bruce1993,Boumali2015,Siouane2024JLTP}
\begin{equation}
\bigl[c^2(p_x+\mathrm{i}m_0\omega x)(p_x-\mathrm{i}m_0\omega x)
     -E^2+m_0^2 c^4\bigr]\,\Psi(x)=0,
\label{eq:KGO1D}
\end{equation}
which has the spectrum
\begin{align}
E_n &= \pm\, m c^2\sqrt{1+2 n r},
\qquad n=0,1,2,\dots, \\
r&\equiv \frac{\hbar\omega}{m c^2}.
\label{eq:spec_KGO}
\end{align}
with unit degeneracy $g_n=1$.  We retain the positive-energy sector
only.  This restriction is fully compatible with the exact
Foldy--Wouthuysen transformation~\cite{Foldy1950}, which decouples
positive- and negative-energy sectors of relativistic-oscillator
Hamiltonians and which has been carried out explicitly for both
the \KGO\ and the \DO\ in Refs.~\cite{Moreno1989,Quesne2017}.  The
mass-shell condition~\eqref{eq:spec_KGO} also follows from the
Feshbach--Villars formulation of the Klein--Gordon equation when
the latter is reduced to a system of two coupled first-order
equations~\cite{Foldy1950}.  The spectrum is bounded from below by
the rest-energy $mc^2$, scales as $\sqrt{n}$ for large $n$ (in
contrast to the linear non-relativistic harmonic spectrum), and
admits an analytic Hurwitz-zeta resummation of its
partition function in the high-$T$
limit~\cite{Boumali2015,Boumali2018DOmin}.

For thermodynamics we do not use the absolute rest-energy shifted
quantity $E_n$ directly.  We use instead the excitation spectrum
\begin{equation}
\epsn \equiv E_n-E_0
       = m c^2\left(\sqrt{1+2nr}-1\right),
\label{eq:excitation}
\end{equation}
so that the ground state has zero excitation energy.  This shift is
thermodynamically harmless for the heat capacity and entropy, but it
is essential numerically: it guarantees $Z\to g_0$, $S\to k_B\ln g_0$
and $C_v\to0$ as $1/\bav\to0$.  Since the 1D ground state is
non-degenerate ($g_0=1$), the low-temperature entropy also tends to
zero.

\subsection{The 1D Dirac oscillator}
\label{subsec:DO}

The 1D~\DO\ is built from the Dirac Hamiltonian by the substitution
$p\to p-\mathrm{i}m_0\omega\beta\hat{x}$, where $\beta$ is the
usual Dirac matrix.  Its spectrum is degenerate with respect to a
discrete spin index $s=\pm 1$~\cite{Pacheco2003,Pacheco2014}:
\begin{equation}
E_{n,s} \;=\; \pm\, m c^2\,\sqrt{1+(2n+1+s)\,r},
\qquad n=0,1,2,\dots
\label{eq:spec_DO_full}
\end{equation}
The $n=0$, $s=-1$ state is the ground state and is non-degenerate;
all other levels appear in spin-degenerate pairs, since the same
energy is reached either with $(n=N, s=-1)$ or with $(n=N-1,
s=+1)$.  Reorganising the levels by the joint label
$N\equiv n+(1+s)/2$, the positive sector becomes
$E_N = m c^2\sqrt{1+2 N r}$ with effective degeneracy
\begin{equation}
g_N \;=\;
\begin{cases}
1, & N=0, \\[2pt]
2, & N\ge 1.
\end{cases}
\label{eq:deg_DO}
\end{equation}
The 1D~\DO\ therefore shares its energy levels with the 1D~\KGO\
but has each excited level twofold degenerate --- a direct
manifestation of the spin-orbit coupling absent from the
Klein--Gordon problem.  This is the structural feature that, as we
will show in Sec.~\ref{sec:results}, lifts the entropy of the
\DO\ above that of the \KGO\ by~$\ln 2$ in the high-temperature
limit and reorganises the temperature dependence of the specific
heat.  We emphasize that the doubling structure
in~\eqref{eq:deg_DO} is robust under a variety of generalizations
of the model, including the introduction of a magnetic
field~\cite{Mandal2010,Frassino2017}, of noncommutativity in phase
space~\cite{Boumali2013,Oliveira2023}, of a minimal
length~\cite{Boumali2018DOmin}, of cosmic-string
backgrounds~\cite{Bakke2018,Bouzenada2024}, and of doubly special
relativity~\cite{Boumali2026KG3D}.  The
high-temperature value $C_v\to 2\kB$ characteristic of the 1D~\DO\
under standard Boltzmann statistics is a direct consequence of
this degeneracy structure~\cite{Pacheco2003}.

% =================================================================
\section{Beck's superstatistics for the 1D oscillators}
\label{sec:beck}
% =================================================================

Following Beck and Cohen~\cite{Beck2003} and the construction of
Ref.~\cite{Siouane2024JLTP}, the superstatistical Boltzmann factor
in the low-energy asymptotic regime $\bav\epsn\lesssim 1$ takes the
universal form
\begin{equation}
B(\epsn) \;\approx\; \mathrm{e}^{-\bav \epsn}\,\Pi_q(\bav \epsn),
\label{eq:Bbeck}
\end{equation}
with $\Pi_q(\bav\epsn)$ the polynomial bracket of
Eq.~\eqref{eq:Pi_def}, and where the third-order coefficient
$g(q)$ takes the values~\eqref{eq:gGamma}--\eqref{eq:gF} for the
three principal PDFs.  The partition function and free energy
follow from the formal mapping of the non-equilibrium
superstatistical system to an equivalent equilibrium one with
mean inverse temperature~$\bav$~\cite{Beck2003,Beck2011}:
\begin{align}
Z(\bav,q) &= \sum_n g_n\,B(\epsn), \\
F(\bav,q) &= -\frac{1}{\bav}\,\ln Z(\bav,q).
\label{eq:Z_beck}
\end{align}
All thermodynamic quantities are subsequently obtained from $Z$
by ordinary partial derivatives with respect to~$\bav$
[Eq.~\eqref{eq:thermo} below].  This prescription has been
employed in numerous applications of superstatistics to
relativistic-oscillator
problems~\cite{Siouane2024JLTP,Siouane2024TMP,Korichi2021,Boumali2025DSR}
and is the natural starting point against which to compare the
hyperstatistical analysis.

It is essential to recognise that Eq.~\eqref{eq:Bbeck} is
intrinsically asymptotic.  The polynomial bracket $\Pi_q(\bav\varepsilon)$
becomes negative when
$\tfrac{q-1}{2}(\bav E)^2 + g(q)(\bav E)^3<-1$, after which
$B(\epsn)$ ceases to be a Boltzmann-like weight and must either be
truncated to zero or reinterpreted.  For the F-distribution in
particular, $g_F(q)$ is large and positive over a substantial
range of $q\in(1,3)$, and the bracket grows above unity rather
than decaying, so that $B(\epsn)$ amplifies excited levels rather
than suppressing them.  Figure~\ref{fig:breakdown}(a) below makes
the breakdown structure of $\Pi_q$ explicit.  Independently of the
specific PDF, the asymptotic expansion is reliable only when
$\bav \epsn \ll 1$ for the dominant levels, i.e.\ in the
high-temperature limit, and ceases to be predictive at moderate
and low temperatures.  This is a structural shortcoming of the
framework that is partially mitigated by higher-order
expansions~\cite{Sattin2018} but cannot be eliminated within the
asymptotic approach itself.  In contrast, the hyperstatistical
Boltzmann factor of Eq.~\eqref{eq:Bq} is exact at all temperatures
inside its convergence domain, as we develop next.

% =================================================================
\section{Hyperstatistics for the 1D oscillators}
\label{sec:hyper}
% =================================================================

In the hyperstatistical framework of
Squillante \emph{et al.}~\cite{Squillante2026}, one considers a
$\gamma$-type distribution of \emph{Boltzmann factors} inside each
domain of the system, rather than a distribution of intensive
parameters as in superstatistics.  Equivalently, the inverse
temperature is allowed to fluctuate \emph{within} each domain
according to a $\gamma$-density, while the system as a whole is
treated as a hyperensemble of such domains.  The Laplace transform
of the $\gamma$-distribution evaluates to a power-law factor,
\begin{align}
&\int_0^{\infty}\!\rd\beta_i\,
\frac{1}{\Gamma(n)}\!\left(\frac{n}{\bavi}\right)^{\!n}
\beta_i^{\,n-1} e^{-n\beta_i/\bavi} e^{-\beta_i E}
\notag\\
&\hspace{2.4cm}=\left(1+\frac{\bavi E}{n}\right)^{-n}.
\label{eq:LT_gamma}
\end{align}
This result is exactly a $q$-exponential when
\begin{equation}
q=1+\frac{1}{n}, \qquad n=\frac{1}{q-1},
\label{eq:qn_relation}
\end{equation}
so the equality between $q$ and $n$ belongs to this particular
Gamma-Laplace transform.  It should not be confused with the
broader convergence domain of the $q$-generalized Gamma function,
$1<q<1+1/n$, where $q$ and $n$ can be treated as phenomenological
parameters inside the admissible interval.  This distinction is
important because it resolves the apparent tension between the
exact transform and the fitting practice used in
Ref.~\cite{Squillante2026}.

After the domain-level transform, hyperstatistics introduces the
closed-form effective $q$-generalized Boltzmann factor
\begin{equation}
\boxed{\, B_q(E) \;=\; \expq\!\bigl(-\bav E\bigr).\,}
\label{eq:Bq_box}
\end{equation}
In the present thermodynamic application, Eq.~\eqref{eq:Bq_box} is
used as an effective domain-averaged statistical weight.  We do not
assume the generally invalid nonlinear identity
\begin{equation}
\int_0^\infty f(\beta)\expq(-\beta E)\,\rd\beta
=\expq(-\bav E).
\label{eq:not_identity}
\end{equation}
Rather, the physical interpretation is that the internal
distribution of ordinary Boltzmann factors generates a
$q$-exponential within each domain and the experimentally accessible
mean $\bav$ characterizes the system as a whole.  The functional
form of $B_q$ is then independent of the particular PDF used to
model the domain ensemble; only the effective scale in the argument
is PDF dependent~\cite{Squillante2026}.  We emphasize three
structural features of Eq.~\eqref{eq:Bq_box} that have direct
consequences for the thermodynamics:
\begin{enumerate}
\item \textit{Universality across PDFs.}  The Uniform, $\gamma$,
      Log-Normal, F and $q$-$\gamma$ distributions all give the
      same $q$-exponential closed form (Table~1
      of~\cite{Squillante2026}).  Hyperstatistics therefore
      eliminates the explicit PDF-dependent coefficient $g(q)$
      that controls the asymptotic superstatistics.
\item \textit{Manifest positivity.}  With the convention
      $\expq(-x)=[1+(q-1)x]^{-1/(q-1)}$ for $q>1$, the
      $q$-exponential is strictly positive and monotonically
      decreasing for all $x\ge0$; it has an algebraic tail and no
      finite cut-off.  Compact support occurs instead for $q<1$.
      Thus the hyperstatistical weight never develops the negative
      regions produced by Beck's truncated polynomial bracket.
\item \textit{Tsallis consistency.}  The hyperstatistical
      Boltzmann factor is the unique form compatible with Tsallis'
      nonadditive $q$-entropy~\eqref{eq:Sq} and preserves the
      concavity of $S_q$~\cite{Tsallis1988,Squillante2026}.
\end{enumerate}

The hyperstatistical partition function for either oscillator
follows immediately from~\eqref{eq:Bq_box},
\begin{equation}
Z_q(\bav) \;=\; \sum_n g_n\,\expq\!\bigl(-\bav \epsn\bigr).
\label{eq:Zq}
\end{equation}
We retain the same canonical thermodynamic
prescription~\eqref{eq:Z_beck} so that the comparison with Beck's
framework is on the same footing.  In contrast
to~\eqref{eq:Bbeck}, the $q$-exponential factor~\eqref{eq:Bq_box}
has a single closed analytical form, and its derivatives with
respect to $\bav$ admit the simple identity
\begin{equation}
\frac{\partial}{\partial\bav}\,\expq\!\bigl(-\bav \epsn\bigr)
\;=\; -\,\epsn\,\bigl[\expq\!\bigl(-\bav \epsn\bigr)\bigr]^{q},
\label{eq:dqexp}
\end{equation}
which propagates to higher orders without polynomial blow-up.  In
particular, the second derivative,
\begin{equation}
\frac{\partial^{2}}{\partial\bav^{2}}\,\expq\!\bigl(-\bav \epsn\bigr)
\;=\; q\, \epsn^{2}\,\bigl[\expq\!\bigl(-\bav \epsn\bigr)\bigr]^{2q-1},
\label{eq:d2qexp}
\end{equation}
is again positive for $q>0$, as required for a thermodynamically
stable system.  Equations~\eqref{eq:dqexp}--\eqref{eq:d2qexp} are
the analytical engine that allows us to express the entropy and
specific heat of the \KGO\ and \DO\ in closed form (up to the
sum over levels) in Sec.~\ref{sec:thermo}.

% =================================================================
\section{Thermodynamic prescription}
\label{sec:thermo}
% =================================================================

For both frameworks we use the standard relations
\begin{align}
U &= -\frac{\partial\ln Z}{\partial\bav}, \\
\frac{S}{\kB} &= \ln Z + \bav U, \\
\frac{C_v}{\kB} &= \bav^{2}\frac{\partial^{2}\ln Z}{\partial\bav^{2}}.
\label{eq:thermo}
\end{align}
Setting $\kB=1$, $m c^2=1$ and $r=1$, all thermodynamic
quantities become functions of $(T,q)$ with $T\equiv 1/\bav$ in
units of the rest-energy $mc^2$.  In all numerical plots we use
$\epsn=E_n-E_0$ rather than the absolute relativistic energy.
This is the standard canonical prescription: adding a constant to
all levels multiplies $Z$ by a factor and shifts $U$, but it cannot
change $C_v$.  Using $\epsn$ makes the third-law limit explicit and
prevents the spurious low-temperature divergence produced by an
unsubtracted rest-energy contribution.  Sums over $n$ in
\eqref{eq:Z_beck} and~\eqref{eq:Zq} are performed numerically,
retaining $N=8000$ levels.  For the hyperstatistical sums the
large-$n$ behaviour $\epsn\sim\sqrt{n}$ implies convergence only
for $q<3/2$; all plots therefore use $q\le1.40$.  For Beck's
framework the derivatives are obtained analytically from the
truncated weight, with a level-by-level positivity cut-off applied
only where the polynomial bracket becomes negative.

For hyperstatistics, the derivative
identities~\eqref{eq:dqexp}--\eqref{eq:d2qexp} allow $U$ and the
second derivative of $\ln Z$ to be expressed analytically in terms
of partition-function-like sums:
\begin{align}
U &= \frac{1}{Z_q}\sum_n g_n\,\epsn\,
        \bigl[\expq(-\bav \epsn)\bigr]^{q}, \label{eq:U_hyper}\\[2pt]
\frac{\partial^2\ln Z_q}{\partial\bav^2}
  &= \frac{q}{Z_q}\sum_n g_n\,\epsn^{2}\,
        \bigl[\expq(-\bav \epsn)\bigr]^{2q-1}
     \;-\; U^{2}.
\label{eq:d2lnZ_hyper}
\end{align}
The combination of Eqs.~\eqref{eq:Zq},
\eqref{eq:U_hyper}--\eqref{eq:d2lnZ_hyper} and~\eqref{eq:thermo}
is the essential improvement of hyperstatistics over Beck's
asymptotic framework at the computational level: thermodynamic
quantities are obtained as level sums of the closed-form
$q$-exponential and its powers, without numerical
differentiation, without smoothing kernels, and without the
positivity cut-off that pollutes Beck's polynomial.  We treat
the 1D~\KGO\ and 1D~\DO\ with the degeneracies of
Sec.~\ref{sec:spectrum}: $g_n=1$ for the \KGO\ and
$g_0=1,\ g_n=2$ for $n\ge 1$ for the~\DO.

% =================================================================
\section{Results and discussion}
\label{sec:results}
% =================================================================

\subsection{Hyperstatistical thermodynamics of the 1D \KGO\ and 1D \DO}
\label{subsec:hyper_results}

\begin{figure*}[t]
  \centering
  \includegraphics[width=\linewidth]{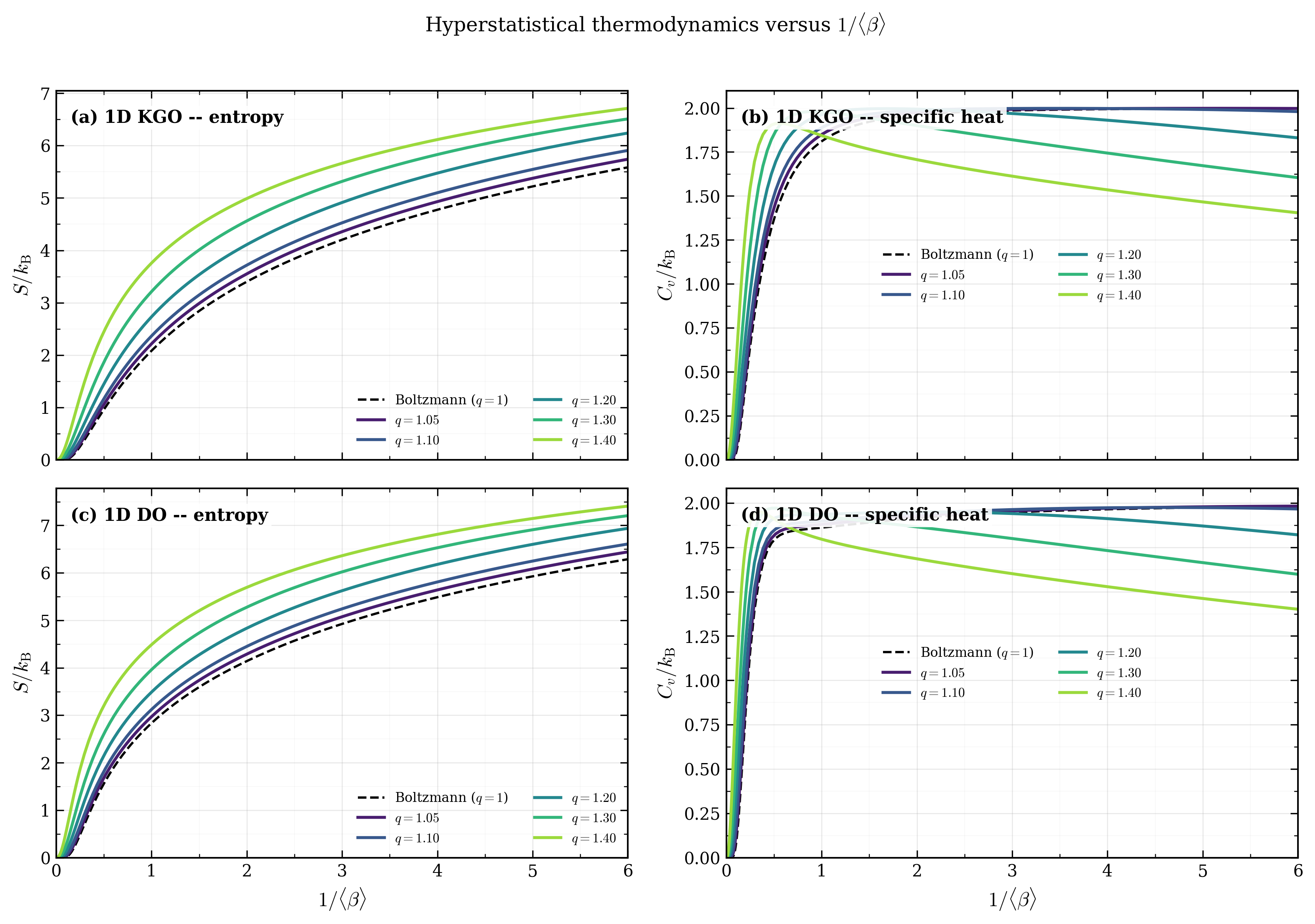}
  \caption{Hyperstatistical thermodynamics of the 1D~\KGO\ (top
  row) and 1D~\DO\ (bottom row) plotted against
  $1/\bav$.  Left column: entropy~$S/\kB$; right column:
  specific heat~$C_v/\kB$.  Each colour corresponds to a value of
  the entropic index~$q$ (legend); the dashed black curve in every
  panel is the Boltzmann limit ($q=1$).  The use of excitation
  energies $\epsn=E_n-E_0$ enforces $S\to0$ and $C_v\to0$ as
  $1/\bav\to0$.  Degeneracies: $g_n=1$ for the \KGO\ and $g_0=1,\
  g_n=2$ ($n\ge 1$) for the \DO.}
  \label{fig:hyper}
\end{figure*}

Figure~\ref{fig:hyper} shows the entropy and specific heat of both
oscillators under hyperstatistics for $q=1.0$ (Boltzmann), $1.05$,
$1.10$, $1.20$, $1.30$ and $1.40$, plotted as functions of the
temperature variable $1/\bav$.  This representation is physically
more transparent than plotting against $\bav$: the left side is the
low-temperature regime and the right side is the high-temperature
regime.  Because the partition function is built from the excitation
energies $\epsn$, the specific heat of both oscillators vanishes
as $1/\bav\to0$.  Several features stand out:
\begin{itemize}
\item In the Boltzmann limit ($q=1$), $C_v$ starts from zero at
      $1/\bav=0$, rises smoothly with thermal activation of the excited
      levels, and approaches the high-temperature value near
      $2\kB$.  The entropy also starts from zero because the ground
      state is non-degenerate.  The \DO\ entropy lies above the
      \KGO\ entropy at high temperature by an increment that
      asymptotes to $\ln 2$, the entropic signature of the
      spin-induced degeneracy doubling discussed in
      Sec.~\ref{subsec:DO}.
\item Increasing $q$ above $1$ broadens the thermal activation region:
      the $q$-exponential has an algebraic tail, so excited states
      are populated more efficiently than in the Boltzmann case.
      Nevertheless, because the zero of energy is the ground state,
      all excited-state weights still vanish as $1/\bav\to0$, and
      $C_v\to0$ for every plotted $q<3/2$.  The approach is slower
      than in Boltzmann statistics, but it is finite and physical.
      These long-tail features are the hyperstatistical analogue of
      the $q$-relaxation phenomenology already documented in many
      condensed-matter and turbulence contexts~\cite{Beck2003,Beck2011,Squillante2026}.
\item The two oscillators differ structurally only in degeneracy:
      the \DO\ exhibits the same qualitative $C_v$ profile as the \KGO\,
      with a small upward shift, and the same temperature-dependent entropy
      curve raised by approximately $\ln 2$ at small $\bav$.  The
      level-by-level positivity of $B_q(E)$ means that no spurious
      oscillation appears anywhere in the displayed range; the
      curves are smooth, monotonic where physically expected, and
      analytic in $q$.
\end{itemize}

\subsection{Comparison with Beck's superstatistics: 1D \KGO}
\label{subsec:compare_KGO}

\begin{figure*}[t]
  \centering
  \includegraphics[width=\linewidth]{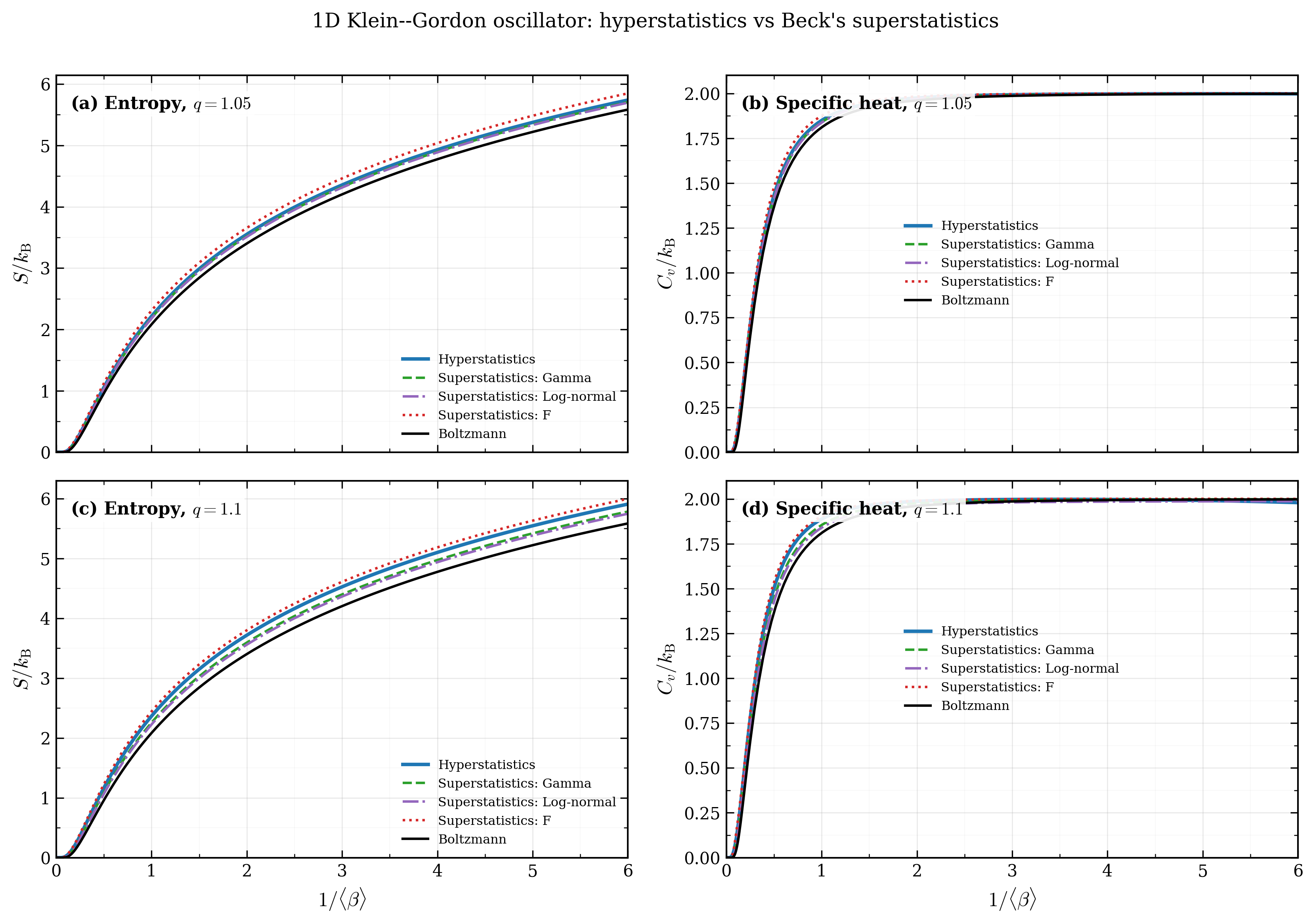}
  \caption{1D Klein--Gordon oscillator: hyperstatistical
  predictions (closed-form $B_q(\varepsilon)=\expq(-\bav\varepsilon)$, solid blue)
  compared with Beck's superstatistics asymptotic
  expansion~\eqref{eq:Bbeck} for the Gamma (dashed green),
  Log-Normal (dot-dashed purple) and F (dotted red) distributions,
  at $q=1.05$ (top row) and $q=1.10$ (bottom row).  Left column:
  entropy~$S/\kB$ vs $1/\bav$; right column: specific heat
  $C_v/\kB$ vs $1/\bav$.  The thin solid black curve is the
  Boltzmann reference $q=1$.  The curves are plotted against
  $1/\bav$.  Beck's polynomial bracket remains meaningful only while $\bav\epsn$ is small enough for the dominant levels; outside this region the asymptotic expansion ceases to be a positive Boltzmann-like weight (see Fig.~\ref{fig:breakdown}).}
  \label{fig:compareKGO}
\end{figure*}

Figure~\ref{fig:compareKGO} compares the hyperstatistical entropy
and specific heat of the 1D~\KGO\ with those obtained from Beck's
polynomial bracket~\eqref{eq:Bbeck} for the Gamma, Log-Normal and
F-distributions, at $q=1.05$ and $q=1.10$.  The comparison is shown directly as a function of
$1/\bav$.  Beck's polynomial bracket is reliable only where the
product $\bav\epsn$ remains moderate for the levels that dominate
$Z$ (see Fig.~\ref{fig:breakdown} for the breakdown beyond this
range).  In the high-temperature regime all four curves agree at
the percent level,
confirming the validity of the polynomial expansion in its
low-energy domain.  As $1/\bav$ decreases toward the low-temperature regime:
\begin{itemize}
\item the Gamma- and Log-Normal-superstatistics specific heats
      track each other closely and approach the Boltzmann curve
      from above; this is the same regularity already noted for
      the higher-dimensional \KGO\
      in~\cite{Siouane2024JLTP}: for these PDFs $g(q)$ is small
      and the cubic correction is sub-leading throughout the
      validity window;
\item the F-superstatistics curve sits noticeably above the
      others, consistent with its larger and positive $g_F(q)$,
      which lifts $B(\epsn)$ at moderate $\bav\epsn$ and pushes more
      weight into excited levels;
\item the hyperstatistical curve (solid blue) interpolates
      between the Gamma/Log-Normal cluster and the F-distribution
      at intermediate temperature, then returns smoothly toward zero at
      low temperature because the sums are built from excitation
      energies.  The curve is smooth in the displayed interval and
      avoids the small numerical distortions that may appear when
      Beck's polynomial bracket approaches its positivity boundary.
\end{itemize}

\subsection{Comparison with Beck's superstatistics: 1D \DO}
\label{subsec:compare_DO}

\begin{figure*}[t]
  \centering
  \includegraphics[width=\linewidth]{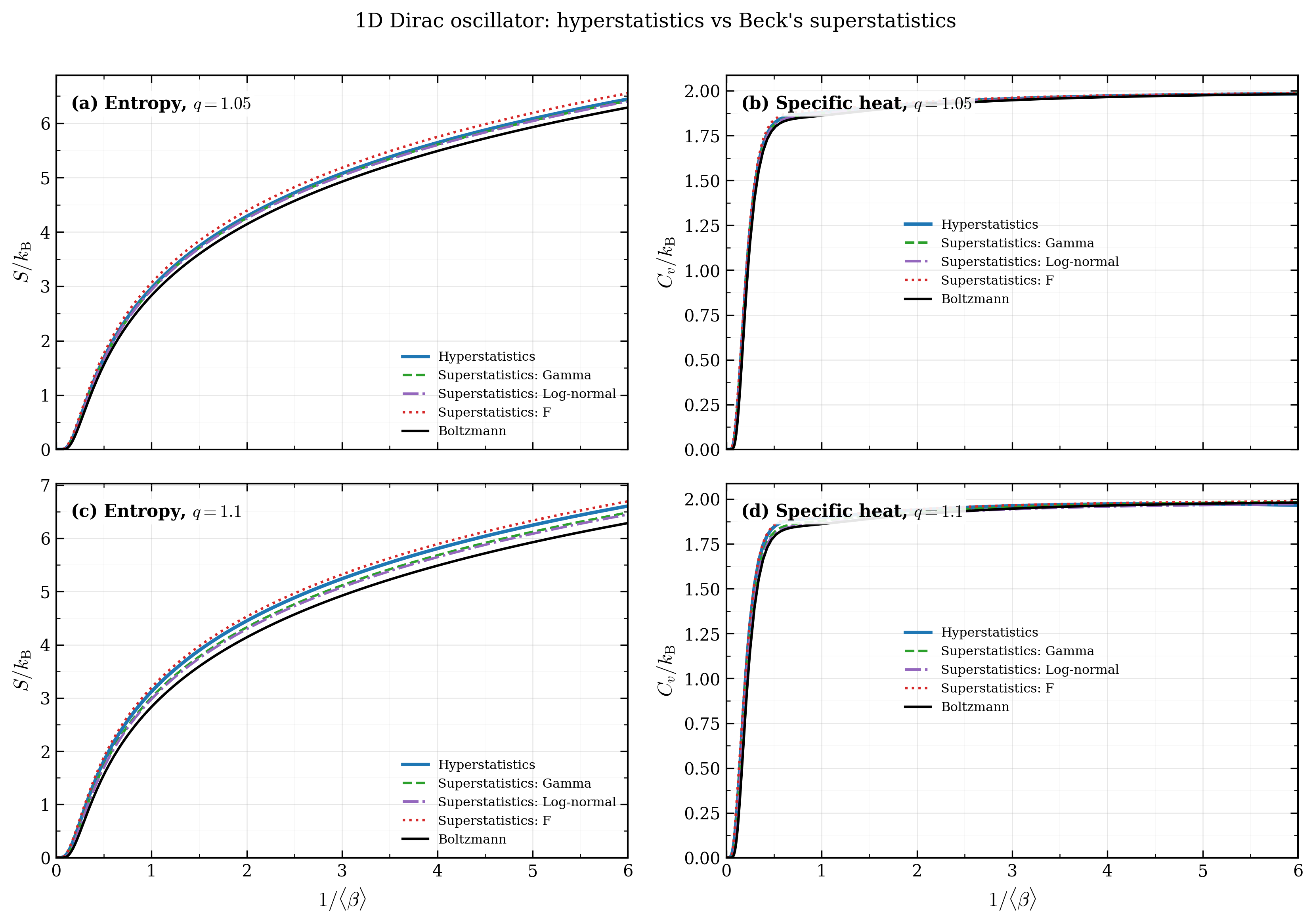}
  \caption{1D Dirac oscillator: same comparison as
  Fig.~\ref{fig:compareKGO} but with the degeneracies $g_0=1$,
  $g_n=2$ ($n\ge 1$).  The doubling of every excited level lifts
  the entropy by $\approx\ln 2$ in the high-temperature limit relative
  to the \KGO\ of Fig.~\ref{fig:compareKGO} and reduces the spread
  of $C_v$ across the four prescriptions, since the partition
  function is dominated by the now-weightier excited levels.  The curves are plotted against $1/\bav$.}
  \label{fig:compareDO}
\end{figure*}

Figure~\ref{fig:compareDO} repeats the comparison for the 1D
Dirac oscillator, with the degeneracy $g_0=1$, $g_n=2$ ($n\ge 1$)
of Eq.~\eqref{eq:deg_DO}.  The qualitative picture is the same as
in Fig.~\ref{fig:compareKGO} but quantitatively shifted:
\begin{itemize}
\item the entropy is uniformly larger than that of the \KGO,
      with the difference asymptoting to $\ln 2$ in the
      high-temperature limit --- the entropic signature
      of the doubled degeneracy of the excited levels;
\item the specific heat curves bunch closer to $2\kB$ across the
      displayed range, reflecting the heavier weighting of
      high-energy levels via $g_n=2$ (which makes the
      partition function dominated by excited-level contributions
      already at moderate temperature);
\item the spread of $C_v$ across the four prescriptions
      (hyperstatistics, Gamma, Log-Normal, F) is reduced compared
      with the \KGO\ case, again because each individual level
      contributes a smaller relative fraction of $Z$ when
      $g_n=2$ for $n\ge 1$;
\item the F-distribution remains the outlier on the high side,
      whereas the hyperstatistical curve remains the smoothest of the four
      in the comparison window.
\end{itemize}

\subsection{\KGO\ vs \DO\ under hyperstatistics and the low-temperature check}
\label{subsec:KGOvsDO}

\begin{figure}[t]
  \centering
  \includegraphics[width=\linewidth]{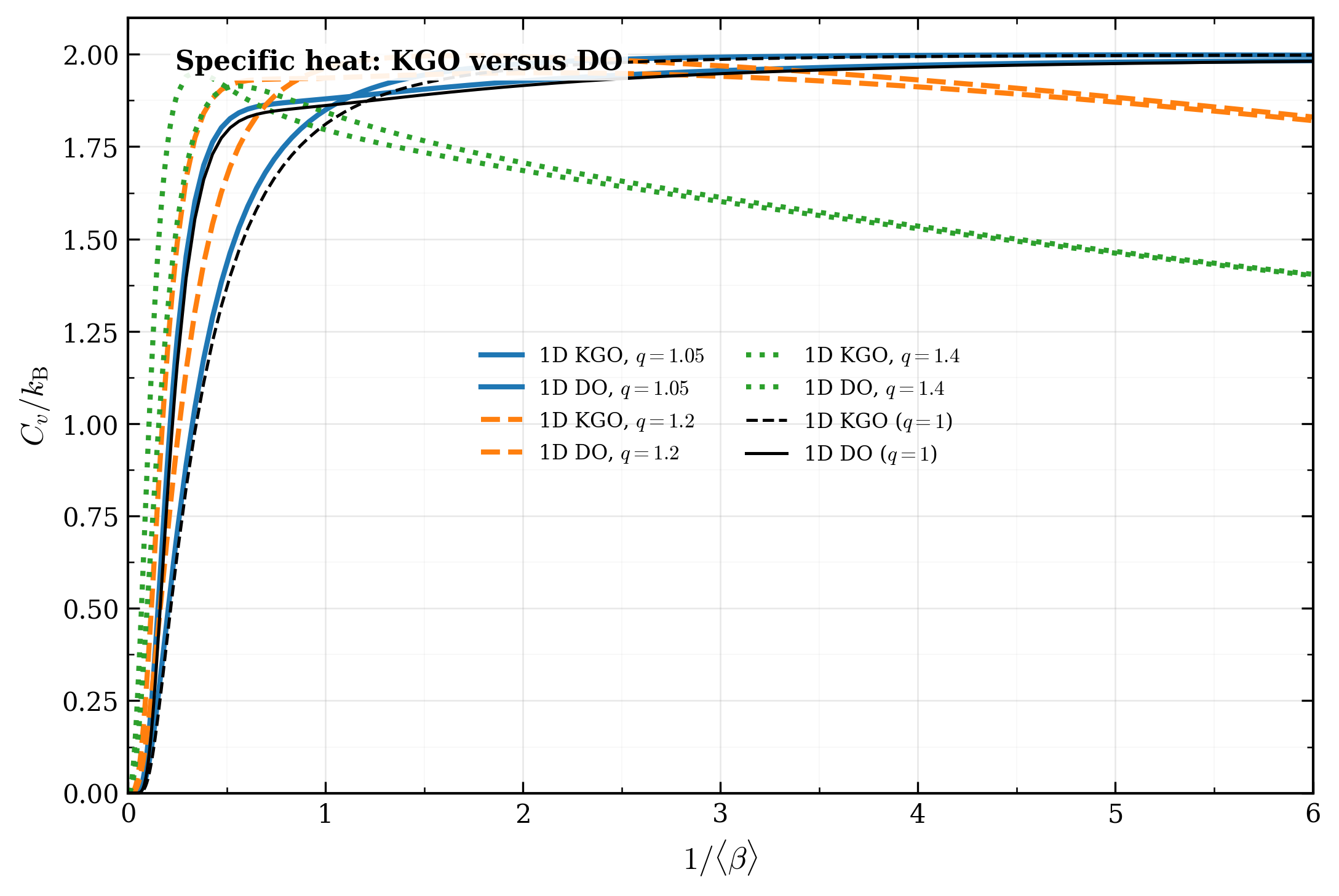}
  \caption{Hyperstatistical specific heat of the 1D~\KGO\ and
  1D~\DO\ versus $1/\bav$ for representative values of $q$.  The
  black curves give the Boltzmann limit.  The excited-state
  degeneracy of the \DO\ lifts the curve relative to the \KGO\ in
  the thermally active region, while both models satisfy
  $C_v\to0$ as $1/\bav\to0$.}
  \label{fig:KGOvsDO}
\end{figure}

Figure~\ref{fig:KGOvsDO} overlays the hyperstatistical specific
heats of the \KGO\ and \DO\ at $q=1.05$, $1.20$ and $1.40$,
together with the Boltzmann limits.  The horizontal axis is
$1/\bav$.  The \DO\ curve is slightly above the \KGO\ curve over
most of the thermally active region because every excited level is
doubled.  At very low temperature, however, both models return to a
non-degenerate ground state and therefore both heat capacities tend
to zero.  This behaviour is the decisive check that the excitation
spectrum $\epsn=E_n-E_0$ has been used consistently.

\begin{figure}[t]
  \centering
  \includegraphics[width=\linewidth]{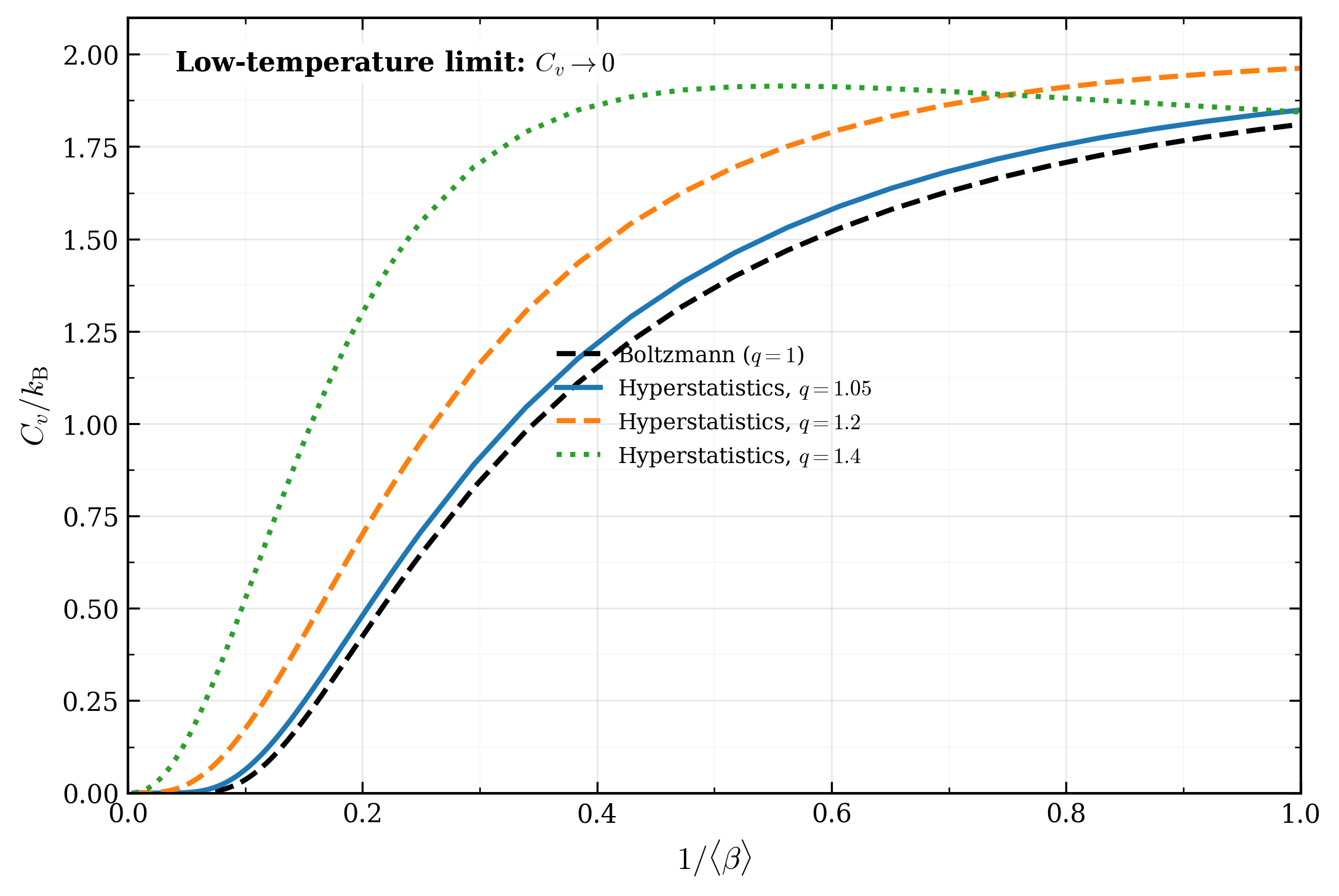}
  \caption{Low-temperature check for the 1D~\KGO.  The specific
  heat tends to zero as $1/\bav\to0$ for Boltzmann statistics and for
  hyperstatistics with $q<3/2$.  This confirms that the numerical
  procedure is free of the unphysical low-temperature divergence.}
  \label{fig:lowT}
\end{figure}

Figure~\ref{fig:lowT} magnifies the low-temperature part of the
\KGO\ heat capacity.  The apparent divergence that can occur when
one inserts the absolute relativistic energies directly in the
thermodynamic sums is absent.  With $\epsn=E_n-E_0$, the ground
state has zero excitation energy and all excited-state weights
vanish as $1/\bav\to0$; consequently $C_v$ approaches zero for the
Boltzmann curve and for the hyperstatistical curves with
$q<3/2$.

\subsection{Where Beck's polynomial bracket breaks down}
\label{subsec:breakdown}

\begin{figure}[t]
  \centering
  \includegraphics[width=\linewidth]{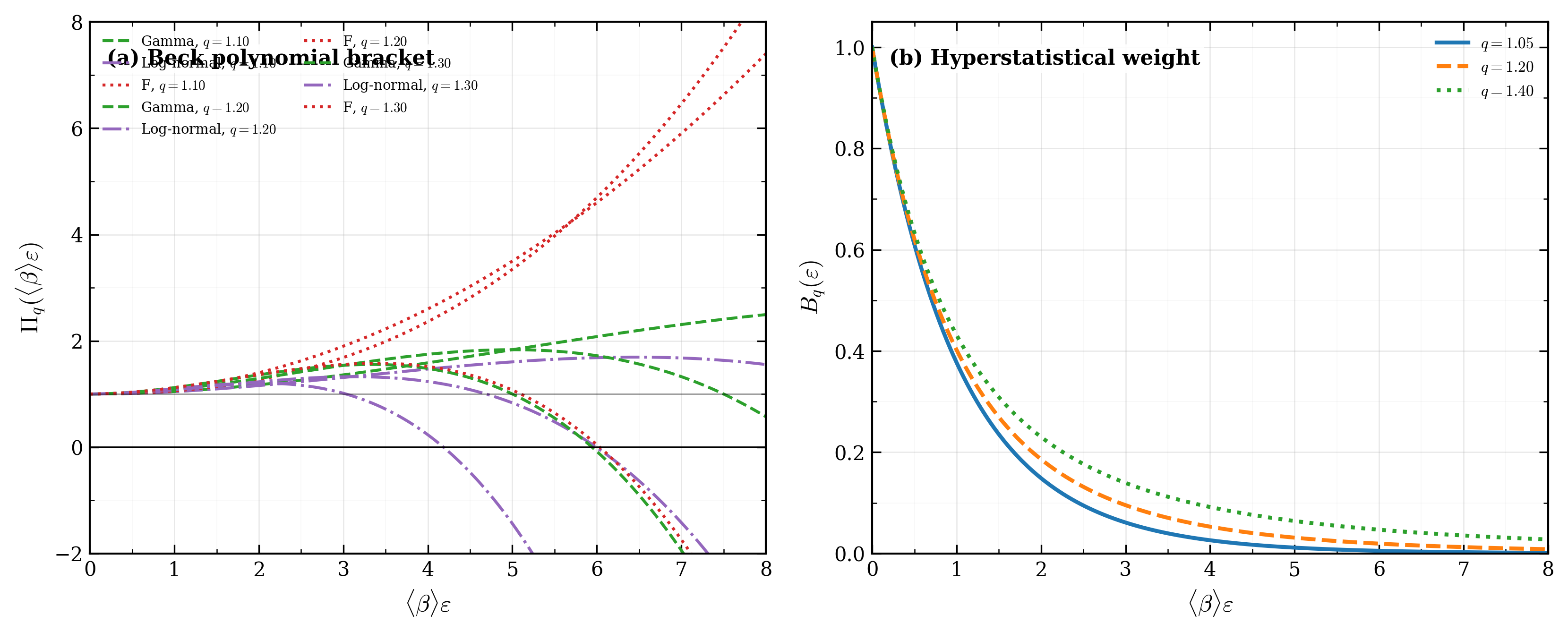}
  \caption{Beck's asymptotic series (panel~a) compared with the
  hyperstatistical generalized Boltzmann factor (panel~b) at
  $\bav=1$.  (a)~The polynomial bracket
  $\Pi_q(\bav\varepsilon)=1+\tfrac{q-1}{2}(\bav\varepsilon)^2 + g(q)(\bav\varepsilon)^3$ for
  the Gamma (green), Log-Normal (purple) and F-distribution
  (red) at $q=1.10$, $1.20$, $1.30$.  The Gamma and Log-Normal
  brackets turn negative for moderate $\bav\varepsilon$ at $q\gtrsim 1.3$,
  while the F-distribution bracket grows without bound, both
  invalidating the asymptotic interpretation of $B(\varepsilon)$ as a
  Boltzmann-like weight.  (b)~The hyperstatistical
  $B_q(\varepsilon)=\expq(-\bav\varepsilon)$ is positive and monotonically
  decreasing for $q>1$ and $\varepsilon\ge0$, with an algebraic tail rather
  than a polynomial artefact.  This breakdown is the key practical
  reason for preferring the hyperstatistical formulation when a
  stable statistical weight is required beyond the small-$\bav\varepsilon$
  expansion.}
  \label{fig:breakdown}
\end{figure}

Figure~\ref{fig:breakdown}(a) plots the polynomial bracket
$\Pi_q(\bav\varepsilon)$ of~\eqref{eq:Pi_def} as a function of $E$ at
$\bav=1$ for the three PDFs and $q=1.10$, $1.20$, $1.30$.  Two
distinct failure modes are visible:
\begin{itemize}
\item For the Log-Normal and Gamma distributions the bracket dips
      below zero around $\bav\varepsilon\approx 5$ at $q=1.30$, with the
      Log-Normal failing slightly earlier than the Gamma.  Once
      $\Pi_q<0$, the corresponding levels carry a formally
      negative weight in the asymptotic representation, which we
      truncate to zero in the numerical computation but which
      cannot be reconciled with the Boltzmann interpretation
      $B(\varepsilon)\ge 0$.
\item For the F-distribution, $\Pi_q$ grows without bound,
      crossing $\Pi_q\approx 2$ already at $\bav\varepsilon\approx 4$ for
      $q=1.10$ and exceeding $6$ at $\bav\varepsilon\approx 7$.  This is
      not an underflow problem but a structural one: $g_F(q)$
      changes sign at $q=6/5$ and remains large and positive over
      a substantial part of the physical range $q\in(1,3)$, and
      the cubic term overwhelms the quadratic at moderate
      energies.
\end{itemize}
By contrast, Fig.~\ref{fig:breakdown}(b) shows that
$\expq(-\bav\varepsilon)$ stays strictly positive and monotonically
decreasing for the physically relevant case $q>1$ and $\varepsilon\ge0$.
There is no upper energy cut-off in this case; the decay is
algebraic, with asymptotic behaviour
$\expq(-\bav\varepsilon)\sim \varepsilon^{-1/(q-1)}$.  This represents the principal
practical advantage of hyperstatistics for the
relativistic-oscillator problem: the asymptotic expansion is
replaced by a closed-form weight that admits no spurious sign
changes and propagates analytically into the thermodynamic
relations~\eqref{eq:U_hyper}--\eqref{eq:d2lnZ_hyper}.  We
emphasize that the breakdown of Fig.~\ref{fig:breakdown}(a) is not
a numerical artefact of the asymptotic series but a structural
feature of a finite polynomial approximation.  If the physical
input is an experimentally known distribution $f(\beta)$, one may
avoid this difficulty in superstatistics by evaluating the full
Laplace transform rather than truncating it.  However, the
resulting function is generally PDF dependent, whereas
hyperstatistics uses the same closed $q$-exponential form after the
domain-level construction.

\subsection{Extended comparison and selection of the most appropriate framework}
\label{subsec:best_method}

The comparison between superstatistics and hyperstatistics should
not be reduced to the numerical value of $R^2$ or to the smoothness
of a single curve.  The two methods answer different modelling
questions.  Superstatistics starts from a fluctuating intensive
parameter and computes the Laplace transform of its PDF; it is
therefore the most physically direct framework when the distribution
of inverse temperature, relaxation rate, or another intensive
quantity is experimentally known.  Hyperstatistics starts one level
higher, by assuming that the local statistical weights themselves
are already non-Boltzmann-Gibbsian at the domain scale; it is
therefore more appropriate when the data show robust power-law
occupation or relaxation and when one wants a closed, positive and
PDF-independent effective weight.

\begin{description}
\item[Fluctuating object.] Beck superstatistics fluctuates an intensive
variable, usually $\beta$, and starts from
$B(\varepsilon)=\int f(\beta)e^{-\beta\varepsilon}\,\rd\beta$.  Hyperstatistics
fluctuates the statistical weights generated inside domains and uses
the effective system weight $B_q(\varepsilon)=\expq(-\bav\varepsilon)$.
\item[Physical picture.] Superstatistics assumes locally
Boltzmann--Gibbs cells with slowly varying temperature or rate.
Hyperstatistics assumes domains that already display
non-Boltzmann--Gibbsian statistics and long-tailed statistical
weights.
\item[Dependence on the PDF.] Superstatistics is strongly PDF
dependent: Gamma, Log-Normal, and F distributions lead to different
analytical or asymptotic expressions.  Hyperstatistics is less sensitive
to this choice: the PDF changes the effective scale entering the
argument, while the final form remains $q$-exponential.
\item[Mathematical status in this paper.] The Beck expression used
here is a low-energy expansion and is reliable only while $\bav E$ is
small and the polynomial bracket remains positive.  The
hyperstatistical weight is a closed effective $q$-exponential and is
positive for $q>1$ and $\varepsilon\ge0$.
\item[Numerical stability.] Beck's truncated expansion requires
smoothing and a positivity cut-off when the bracket fails.
Hyperstatistics provides analytical derivatives and avoids negative
statistical weights.
\item[Best use.] Beck superstatistics is most suitable when the measured
fluctuations of $\beta$ or of a relaxation rate are the central
physics, especially if the full Laplace transform can be evaluated.
Hyperstatistics is most suitable when one needs a compact positive model for
thermodynamic sums over broad energy ranges, as in the present
oscillator analysis.
\end{description}

For the present one-dimensional \KGO/\DO{} thermodynamics, the most
appropriate practical framework is hyperstatistics.  This does not mean
that superstatistics is intrinsically invalid, but that the version usually used for
these oscillator applications is the truncated Beck expansion.  That
expansion is reliable only for $q-1\ll1$ and $\bav E\lesssim2$;
outside this window it may produce negative or increasing weights.
Hyperstatistics avoids these artefacts and gives smoother entropy
and heat-capacity curves while requiring fewer ad hoc numerical choices.
If future experiments or simulations provide a measured distribution
$f(\beta)$ for the oscillator bath, the most appropriate strategy would be to
use the exact superstatistical Laplace transform as a benchmark and
then compare it with the hyperstatistical $q$-exponential.  In the
absence of such measured microscopic fluctuations, hyperstatistics is
the cleaner and more robust working framework for the present paper.

% =================================================================
\section{Conclusions}
\label{sec:conclusion}
% =================================================================

We have rewritten the superstatistical analysis of the 1D
Klein--Gordon oscillator~\cite{Siouane2024JLTP} within the
hyperstatistical framework of
Squillante \emph{et al.}~\cite{Squillante2026} and extended it to
the 1D Dirac oscillator with its appropriate spin-induced
degeneracy.  The closed form $B_q(\varepsilon)=\expq(-\bav\varepsilon)$ ---
independent of the chosen probability density $f(\beta)$ ---
leads to a partition function and to thermodynamic functions that
admit clean analytical derivatives via the
$q$-exponential identity~\eqref{eq:dqexp} and that are smooth in
$1/\bav$ throughout the convergence domain of the
$q$-exponential.  The use of excitation energies
$\epsn=E_n-E_0$ also guarantees the physically required limit
$C_v\to0$ as $1/\bav\to0$.

A direct comparison with Beck's low-energy polynomial expansion
shows that the two frameworks coincide quantitatively for
$q-1\ll 1$ and $\bav\varepsilon\lesssim 2$, but that
the polynomial bracket loses positivity at moderate $\bav\varepsilon$ and
$q\gtrsim 1.2$ for the Gamma and Log-Normal PDFs and grows
unboundedly for the F-distribution at the same $\bav\varepsilon$.  These
breakdowns produce spurious oscillations or sharp dips in the
asymptotic-superstatistical thermodynamic functions when the
third-order term in $\Pi_q$ overtakes the second.  The
hyperstatistical curves, by contrast, are smooth in $1/\bav$ at all
temperatures and approach the Boltzmann limit $C_v\to 2\kB$ in
the high-$T$ (small-$\bav$) regime for $q=1$, in agreement with
Refs.~\cite{Pacheco2003,Boumali2015,Siouane2024JLTP,Siouane2024TMP}.

The structural difference between the two oscillators is small in
$C_v$ but conspicuous in $S$, where the doubled degeneracy of the
\DO\ lifts the entropy by $\ln 2$ in the high-temperature limit
relative to the \KGO\ at every~$q$.  This entropic signature is a
direct manifestation of the spin-orbit coupling that distinguishes
the Dirac oscillator from its Klein--Gordon counterpart, and it
is reproduced by both frameworks within their common region of
validity.  The hyperstatistical reformulation thus provides a
unified, PDF-independent description of the thermodynamics of
these relativistic oscillators that is both numerically stable and
analytically tractable.  The extended comparison above shows that
hyperstatistics is the best practical method for the present problem, while Beck
superstatistics remains the best physically grounded method when the
underlying distribution of intensive variables is known and the full
Laplace transform can be used.

Several extensions of this work are within reach.  First, the same
analysis can be carried out in 2D and 3D, where the degeneracy
structure is richer and the partition function admits an
Euler--Maclaurin or Hurwitz-zeta closure analogous to the
treatments of~\cite{Boumali2015,Pacheco2014,Siouane2024JLTP}.
Second, the inclusion of an external magnetic
field~\cite{Mandal2010,Frassino2017,Boumali2015graphene} or of a
cosmic-string topological background~\cite{Bouzenada2024} would
test the robustness of the hyperstatistical Boltzmann factor in
the presence of additional non-trivial level structure.  Third,
the recent doubly-special-relativity extensions of the \KGO\ and
\DO~\cite{Boumali2025DSR,Boumali2026KG3D} provide a natural
laboratory in which to study how Planck-scale corrections to the
spectrum interact with the long-tail behaviour of the
$q$-exponential.  Fourth, hyperstatistics has been shown
to regularize the divergences of response functions at critical
points~\cite{Soares2025Gruneisen,Soares2026Schmidt}; an analogous
analysis for the relativistic oscillators in confining or
fractional backgrounds~\cite{Korichi2021} would tie the present
results to the broader programme of nonextensive critical
phenomena.  We leave these directions for future work.

\begin{acknowledgments}
A.B. acknowledges the support of the Laboratory of Theoretical
and Applied Physics at Echahid Cheikh Larbi Tebessi University,
Tebessa.
\end{acknowledgments}

% =================================================================
% References
% =================================================================


\begin{thebibliography}{99}\setlength{\itemsep}{2pt}

\bibitem{Bruce1993}
S.~Bruce and P.~Minning,
``The Klein--Gordon oscillator,''
Nuovo Cimento A \textbf{106}, 711 (1993).
\href{https://doi.org/10.1007/BF02787240}
     {DOI:~10.1007/BF02787240}.

\bibitem{Boumali2015}
A.~Boumali,
``Thermal properties of the one-dimensional Duffin--Kemmer--Petiau
oscillator using Hurwitz zeta function,''
Z.\ Naturforsch.\ A \textbf{70}, 867 (2015).
\href{https://doi.org/10.1515/zna-2015-0140}
     {DOI:~10.1515/zna-2015-0140}.

\bibitem{Moshinsky1989}
M.~Moshinsky and A.~Szczepaniak,
``The Dirac oscillator,''
J.\ Phys.\ A: Math.\ Gen.\ \textbf{22}, L817 (1989).
\href{https://doi.org/10.1088/0305-4470/22/17/002}
     {DOI:~10.1088/0305-4470/22/17/002}.

\bibitem{Ito1967}
D.~It\^o, K.~Mori, and E.~Carriere,
``An example of dynamical systems with linear trajectory,''
Nuovo Cimento A \textbf{51}, 1119 (1967).
\href{https://doi.org/10.1007/BF02721775}
     {DOI:~10.1007/BF02721775}.

\bibitem{Pacheco2003}
M.~H.~Pacheco, R.~R.~Landim, and C.~A.~S.~Almeida,
``One-dimensional Dirac oscillator in a thermal bath,''
Phys.\ Lett.\ A \textbf{311}, 93 (2003).
\href{https://doi.org/10.1016/S0375-9601(03)00467-5}
     {DOI:~10.1016/S0375-9601(03)00467-5}.

\bibitem{Pacheco2014}
M.~H.~Pacheco, R.~V.~Maluf, C.~A.~S.~Almeida, and R.~R.~Landim,
``Three-dimensional Dirac oscillator in a thermal bath,''
EPL \textbf{108}, 10005 (2014).
\href{https://doi.org/10.1209/0295-5075/108/10005}
     {DOI:~10.1209/0295-5075/108/10005}.

\bibitem{Bermudez2007}
A.~Bermudez, M.~A.~Martin-Delgado, and E.~Solano,
``Exact mapping of the Dirac oscillator onto the Jaynes--Cummings
model: Ion-trap experimental proposal,''
Phys.\ Rev.\ A \textbf{76}, 041801(R) (2007).
\href{https://doi.org/10.1103/PhysRevA.76.041801}
     {DOI:~10.1103/PhysRevA.76.041801}.

\bibitem{Lamata2011}
L.~Lamata, J.~Casanova, R.~Gerritsma, C.~F.~Roos,
J.~J.~Garc\'\i a-Ripoll, and E.~Solano,
``Relativistic quantum mechanics with trapped ions,''
New J.\ Phys.\ \textbf{13}, 095003 (2011).
\href{https://doi.org/10.1088/1367-2630/13/9/095003}
     {DOI:~10.1088/1367-2630/13/9/095003}.

\bibitem{Blatt2012}
R.~Blatt and C.~F.~Roos,
``Quantum simulations with trapped ions,''
Nat.\ Phys.\ \textbf{8}, 277 (2012).
\href{https://doi.org/10.1038/nphys2252}
     {DOI:~10.1038/nphys2252}.

\bibitem{Franco2013}
J.~A.~Franco-Villafa\~ne, E.~Sadurn\'\i, S.~Barkhofen, U.~Kuhl,
F.~Mortessagne, and T.~H.~Seligman,
``First experimental realization of the Dirac oscillator,''
Phys.\ Rev.\ Lett.\ \textbf{111}, 170405 (2013).
\href{https://doi.org/10.1103/PhysRevLett.111.170405}
     {DOI:~10.1103/PhysRevLett.111.170405}.

\bibitem{Boumali2015graphene}
A.~Boumali,
``Thermodynamic properties of the graphene in a magnetic field
via the two-dimensional Dirac oscillator,''
Phys.\ Scr.\ \textbf{90}, 045702 (2015).
\href{https://doi.org/10.1088/0031-8949/90/4/045702}
     {DOI:~10.1088/0031-8949/90/4/045702}.

\bibitem{Beck2003}
C.~Beck and E.~G.~D.~Cohen,
``Superstatistics,''
Physica A \textbf{322}, 267 (2003).
\href{https://doi.org/10.1016/S0378-4371(03)00019-0}
     {DOI:~10.1016/S0378-4371(03)00019-0}.

\bibitem{Beck2011}
C.~Beck,
``Generalised information and entropy measures in physics,''
Phil.\ Trans.\ R.\ Soc.\ A \textbf{369}, 453 (2011).
\href{https://doi.org/10.1098/rsta.2010.0280}
     {DOI:~10.1098/rsta.2010.0280}.

\bibitem{Tsallis1988}
C.~Tsallis,
``Possible generalization of Boltzmann--Gibbs statistics,''
J.\ Stat.\ Phys.\ \textbf{52}, 479 (1988);
\emph{Introduction to Nonextensive Statistical Mechanics:
Approaching a Complex World} (Springer, New York, 2009).
\href{https://doi.org/10.1007/BF01016429}
     {DOI:~10.1007/BF01016429};
\href{https://doi.org/10.1007/978-0-387-85359-8}
     {DOI:~10.1007/978-0-387-85359-8}.

\bibitem{Squillante2026}
L.~Squillante, S.~M.~Soares, C.~Tsallis, and M.~de~Souza,
``Hyperstatistics,''
arXiv:2604.24783 (2026).
\href{https://doi.org/10.48550/arXiv.2604.24783}
     {DOI:~10.48550/arXiv.2604.24783}.

\bibitem{Foldy1950}
L.~L.~Foldy and S.~A.~Wouthuysen,
``On the Dirac theory of spin~$1/2$ particles and its
non-relativistic limit,''
Phys.\ Rev.\ \textbf{78}, 29 (1950).
\href{https://doi.org/10.1103/PhysRev.78.29}
     {DOI:~10.1103/PhysRev.78.29}.

\bibitem{Moreno1989}
M.~Moreno and A.~Zentella,
``Covariance, CPT and the Foldy--Wouthuysen transformation for
the Dirac oscillator,''
J.\ Phys.\ A: Math.\ Gen.\ \textbf{22}, L821 (1989).
\href{https://doi.org/10.1088/0305-4470/22/17/003}
     {DOI:~10.1088/0305-4470/22/17/003}.

\bibitem{Quesne2017}
C.~Quesne,
``The Dirac oscillator: from theory to experiment,''
Mod.\ Phys.\ Lett.\ A \textbf{32}, 1730028 (2017).
\href{https://doi.org/10.1142/S0217732317300282}
     {DOI:~10.1142/S0217732317300282}.

\bibitem{Mandal2010}
B.~P.~Mandal and S.~Verma,
``Dirac oscillator in an external magnetic field,''
Phys.\ Lett.\ A \textbf{374}, 1021 (2010).
\href{https://doi.org/10.1016/j.physleta.2009.12.048}
     {DOI:~10.1016/j.physleta.2009.12.048}.

\bibitem{Frassino2017}
A.~M.~Frassino, D.~Marinelli, O.~Panella, and P.~Roy,
``Thermodynamics of quantum phase transitions of a Dirac
oscillator in a homogenous magnetic field,''
J.\ Phys.\ A: Math.\ Theor.\ \textbf{53}, 185204 (2020).
\href{https://doi.org/10.1088/1751-8121/ab7c1f}
     {DOI:~10.1088/1751-8121/ab7c1f}.

\bibitem{Boumali2013}
A.~Boumali and H.~Hassanabadi,
``The thermal properties of a two-dimensional Dirac oscillator
under an external magnetic field,''
Eur.\ Phys.\ J.\ Plus \textbf{128}, 124 (2013).
\href{https://doi.org/10.1140/epjp/i2013-13124-y}
     {DOI:~10.1140/epjp/i2013-13124-y}.

\bibitem{Oliveira2023}
R.~R.~S.~Oliveira and R.~R.~Landim,
``Thermodynamic properties of the noncommutative Dirac oscillator
with a permanent electric dipole moment,''
Eur.\ Phys.\ J.\ Plus \textbf{138}, 74 (2023).
\href{https://doi.org/10.1140/epjp/s13360-023-03700-3}
     {DOI:~10.1140/epjp/s13360-023-03700-3}.

\bibitem{Boumali2018DOmin}
A.~Boumali, L.~Chetouani, and H.~Hassanabadi,
``Effects of a minimal length on the thermal properties of a
Dirac oscillator,''
Can.\ J.\ Phys.\ \textbf{94}, 1019 (2016).
\href{https://doi.org/10.1139/cjp-2016-0257}
     {DOI:~10.1139/cjp-2016-0257}.

\bibitem{Bakke2018}
K.~Bakke and H.~Mota,
``Dirac oscillator in the cosmic string spacetime in the context
of gravity's rainbow,''
Eur.\ Phys.\ J.\ Plus \textbf{133}, 409 (2018).
\href{https://doi.org/10.1140/epjp/i2018-12273-9}
     {DOI:~10.1140/epjp/i2018-12273-9}.

\bibitem{Bouzenada2024}
A.~Bouzenada, A.~Boumali, R.~L.~L.~Vit\'oria, and C.~Furtado,
``Dynamics of a Klein--Gordon oscillator in the presence of a
cosmic string in the Som--Raychaudhuri space-time,''
Theor.\ Math.\ Phys.\ \textbf{221}, 2193 (2024).
\href{https://doi.org/10.1134/S0040577924120134}
     {DOI:~10.1134/S0040577924120134}.

\bibitem{Korichi2021}
N.~Korichi, A.~Boumali, and Y.~Chargui,
``Statistical properties of the 1D space fractional
Klein--Gordon oscillator,''
J.\ Low Temp.\ Phys.\ \textbf{206}, 32 (2022).
\href{https://doi.org/10.1007/s10909-021-02638-z}
     {DOI:~10.1007/s10909-021-02638-z}.

\bibitem{Boumali2025DSR}
A.~Boumali, N.~Jafari, B.~Shukirgaliyev, and F.~Serdouk,
``Thermal properties of Klein--Gordon oscillator in the context
of Amelino--Camelia and Magueijo--Smolin doubly special
relativity frameworks,''
arXiv:2511.11709 (2025).
\href{https://doi.org/10.48550/arXiv.2511.11709}
     {DOI:~10.48550/arXiv.2511.11709}.

\bibitem{Boumali2026KG3D}
A.~Boumali and N.~Jafari,
``Three-dimensional modified Klein--Gordon oscillator in standard
and generalized doubly special relativity,''
arXiv:2602.22444 (2026).
\href{https://doi.org/10.48550/arXiv.2602.22444}
     {DOI:~10.48550/arXiv.2602.22444}.

\bibitem{Siouane2024JLTP}
S.~Siouane and A.~Boumali,
``On the superstatistical properties of the Klein--Gordon
oscillator using Gamma, log, and F distributions,''
J.\ Low Temp.\ Phys.\ \textbf{217}, 598 (2024).
\href{https://doi.org/10.1007/s10909-024-03222-x}
     {DOI:~10.1007/s10909-024-03222-x}.

\bibitem{Siouane2024TMP}
S.~Siouane, A.~Boumali, and A.~Guvendi,
``Superstatistical properties of the Dirac oscillator with
Gamma, lognormal, and F distributions,''
Theor.\ Math.\ Phys.\ \textbf{219}, 673 (2024).
\href{https://doi.org/10.1134/S0040577924050015}
     {DOI:~10.1134/S0040577924050015}.

\bibitem{Sattin2018}
F.~Sattin,
``Superstatistics and temperature fluctuations,''
Physica A \textbf{530}, 121566 (2019);
arXiv:1804.06359 (2018).
\href{https://doi.org/10.1016/j.physa.2019.121566}
     {DOI:~10.1016/j.physa.2019.121566}.

\bibitem{Soares2025Gruneisen}
S.~M.~Soares, L.~Squillante, H.~S.~Lima, C.~Tsallis, and
M.~de~Souza,
``Universally non-diverging Gr\"uneisen parameter at critical
points,''
Phys.\ Rev.\ B \textbf{111}, L060409 (2025).
\href{https://doi.org/10.1103/PhysRevB.111.L060409}
     {DOI:~10.1103/PhysRevB.111.L060409}.

\bibitem{Soares2026Schmidt}
S.~M.~Soares, L.~Squillante, H.~S.~Lima, C.~Tsallis, and
M.~de~Souza,
``Universal and non-universal facets of quantum critical
phenomena unveiled along the Schmidt decomposition theorem,''
arXiv:2512.11093 (2025).
\href{https://doi.org/10.48550/arXiv.2512.11093}
     {DOI:~10.48550/arXiv.2512.11093}.

\end{thebibliography}
\end{document}